\documentclass[12pt,a4paper,english]{article}
  \pdfoutput=1
\usepackage{lmodern}
\usepackage{hyperref}

\usepackage{ulem}
\usepackage[utf8]{inputenc}
\usepackage{geometry}
\usepackage{color}
\usepackage{float}  
\usepackage{amsmath}
\usepackage{amssymb}
\usepackage{graphicx}
\usepackage{setspace}
\usepackage{subfigure}
\usepackage{amsthm}
\usepackage{amsfonts}
\usepackage{braket}

\makeatletter

\usepackage{slashed}
\usepackage{cite}
\date{}

\makeatother

\usepackage{babel}

\begin{document}

\title{Holographic Subregion Complexity in General Vaidya Geometry}

 \author{Yi Ling$^{1,2}$\thanks{lingy@ihep.ac.cn}, Yuxuan Liu$^{1,2}$\thanks{liuyuxuan@ihep.ac.cn}, Chao Niu$^{3}$\thanks{niuchaophy@gmail.com}, Yikang Xiao$^{1,2}$\thanks{ykxiao@ihep.ac.cn}, Cheng-Yong
 Zhang$^{3}$\thanks{zhangcy@email.jnu.edu.cn}}
\maketitle
\begin{center}
\textsl{$^{1}$Institute of High Energy Physics, Chinese Academy of Sciences, Beijing 100049, China}\\
\textsl{$^{2}$School of Physics, University of Chinese Academy of Sciences,
Beijing 100049, China}\\
\textsl{$^{3}$Department of Physics and Siyuan Laboratory, Jinan University, Guangzhou 510632, China}
\par\end{center}
\begin{abstract}
We investigate general features of the evolution of holographic
subregion complexity (HSC) on Vaidya-AdS metric with a general
form. The spacetime is dual to a sudden quench process in quantum
system and HSC is a measure of the ``difference'' between two
mixed states. Based on the subregion CV (Complexity equals Volume)
conjecture and in the large size limit, we extract out three
distinct stages during the evolution of HSC: the stage of linear
growth at the early time, the stage of linear growth with a
slightly small rate during the intermediate time and the stage
of linear decrease at the late time.
The growth rates of the first two stages are compared
with the Lloyd bound. We find that with some choices of certain
parameter, the Lloyd bound is always saturated at the early time,
while at the intermediate stage, the growth rate is always less
than the Lloyd bound. Moreover, the fact that the behavior of CV conjecture and its version of the subregion in Vaidya spacetime implies that they are different even in the large size limit.
\end{abstract}

\newpage
\tableofcontents{}

\section{Introduction}
The relation between quantum information theory and black hole
physics has been investigated for a long time. Traditionally,
computational complexity is a vital concept in quantum information
theory to measure how hard it is to transfer a quantum state to
another. Remarkably, recent progress reveals that the complexity
can also be applied to describe the growth of the interior of
black holes, thus leading to a novel description of complexity
which is named as holographic complexity
\cite{Stanford:2014jda,Brown:2015bva}.

Recently, the notion of holographic complexity has been
generalized to subregion \cite{Alishahiha:2015rta,Carmi:2016wjl,Bakhshaei:2017qud,Lezgi:2019fqu,Bhattacharya:2019zkb,Zhou:2019jlh,Zhang:2019vgl,Fujita:2018xkl,Karar:2019wjb,Auzzi:2019fnp,Ghosh:2019jgd,Auzzi:2019mah,Braccia:2019xxi},
called holographic subregion complexity (HSC). According to AdS/CFT
correspondence, HSC can describe the difference between two
mixed states in boundary theory \cite{Agon:2018zso} and may be
evaluated by two different conjectures, namely, subregion CV
(Complexity$=$Volume) and CA (Complexity$=$Action) conjectures.
Specifically, for a given boundary region $\mathcal{A}$ on a time
slice, one can construct the corresponding entanglement wedge
$W_1$ as well as the Wheeler-DeWitt (WDW) patch $W_2$. Then
the subregion CA conjecture states that the complexity of a
boundary mixed state (which corresponds to the subregion
$\mathcal{A}$), with respect to some reference state, is given
by the action of the intersection $W_1\cap W_2$
\cite{Carmi:2016wjl,Alishahiha:2018lfv,Caceres:2018luq}. While
the subregion CV conjecture states that the complexity could be
given by the volume of an extremal hypersurface
$\Gamma_{\mathcal{A}}$, which is enclosed by the subregion
$\mathcal{A}$ and the corresponding Hubeny-Rangamani-Takayanagi
(HRT) surface $\gamma_{\mathcal{A}}$
\cite{Ben-Ami:2016qex,Abt:2018ywl,Du:2018uua,Abt:2017pmf,Roy:2017uar,Banerjee:2017qti,Zangeneh:2017tub,Bhattacharya:2018oeq,Zhang:2018qnt,Roy:2017kha}.
Then the equation is expressed by
\begin{equation}
  C_{\mathcal{A}}=\frac{V(\Gamma_{\mathcal{A}})}{R\,G_{N}}
\end{equation}
where $R$ is some length scale associated with the geometry.

An intriguing topic is to investigate the evolution behavior of HSC during a holographic quench process, which may be described by the Vaidya-AdS
spacetime \cite{Chen:2018mcc,Ling:2018xpc}. During the quench
 process, a null shell collapses from
the AdS boundary and finally forms an AdS black hole. Finding the
extremal surface $\Gamma_{\mathcal{A}}$ is generally difficult in
these cases. Since the time-reflection symmetry is broken, the
extremal surface $\Gamma_{\mathcal{A}}$ does not simply lie in
a time slice. Fortunately, when the spacetime preserves the
translational symmetry in some spatial directions,
$\Gamma_{\mathcal{A}}$ takes a simple form as shown in the Sec.
\ref{sec_V}.

In the $(2+1)$-dimensional Schwarzschild-AdS (SAdS) case, when the
size of strip $\mathcal{A}$ is far greater than the radius of
horizon, HSC grows linearly and then decreases continuously to
equilibrium. While in higher dimensions, the evolution exhibits a
discontinuous drop if the size of strip $\mathcal{A}$ is large
\cite{Chen:2018mcc}. The above numerical analysis is quite
intuitive but also limited. In particular, the size of strip
$\mathcal{A}$ can not be arbitrarily large in order to keep the numerical
simulations under control. The limitation in numerics
prevents us from exploring the universal behavior of HSC at late
time while it is vital.

In addition, in trying to understand the physical
significance of HSC, it is natural to ask what is the
difference between the CV conjecture and
its version of the subregion, and what makes them different. A
naive understanding is that when the subregion is large enough
to cover the whole boundary, the Hilbert spaces of these two
regions should coincide, thus two conjectures become the
same. That is to say, in large size limit, the growth rate of
HSC should be in agreement with the result of CV conjecture.
However, the limitation of the numerical computation prevents
us from checking the consistency of these two conjectures. Thus a
new approach is needed to investigate this issue.

We further notice that in the thin-shell limit, which means the
thickness of the null shell vanishes, the whole dynamical process
can be described analytically. Therefore, in this paper, we intend to
investigate the evolution of HSC over Vaidya-AdS spacetime by
means of an analytical approach. The strategy that we adopt is
quite in parallel to that as proposed in \cite{Liu:2013qca}, in
which the behavior of entanglement entropy for some non-local
probes is described during the course of a holographic quench. As
shown in Sec.\ref{sec_GVG}, the Vaidya geometry we consider here
is quite general, applicable to different kinds of black hole
backgrounds such as SAdS black hole, Reissner-Nordstrom-AdS
(RN-AdS) black hole and so on. Our work indicates that the HSC
demonstrates similar dynamical behavior in a series of quantum
systems. Specifically, for a strip subregion $\mathcal{A}$, we
analytically investigate the growth behavior of the corresponding
HSC. At the very early time, the HSC increases linearly no matter
what size the subregion is. At the intermediate stage, when the
subregion is sufficiently large, the HSC grows linearly as well.
At the late time stage, if the transition is discontinuous, the HSC increases linearly
as well, otherwise, the HSC will decrease linearly. In addition,
we compare the CV conjecture with its version of the subregion in
the large size limit. We find the results based on two conjectures are different, even in the large scale limit.

The paper is organized as follows: In section \ref{sec_TS}, we
firstly present the setup of general Vaidya geometry and consider a
strip as the subregion $\mathcal{A}$ on the boundary. Next, we derive
some solutions for the corresponding HRT surface and draw the
general configurations of HRT surfaces at the intermediate stage.
Finally, we derive the expression of HSC following the subregion
CV conjecture. In section \ref{sec_GCS}, we characterize the
evolution of HSC as three distinct stages and compare the growth
rate of HSC with the Lloyd bound. In section \ref{sec_Con}, we
show the main conclusions and some discussions.

\section{The Setup}\label{sec_TS}
In this section, we will firstly introduce the general Vaidya
metric in the form of Eddington-Finkelstein coordinates. Then for
a strip boundary region $\mathcal{A}$, we will obtain some
solutions for the corresponding HRT surface which are crucial for
the derivation of the HSC. After that we turn to describe the
configuration of critical HRT surfaces which is essential for us
to understand the behavior of HSC during the intermediate stage of
the evolution. In the end of this section, we will derive the
expression of HSC.
\subsection{General Vaidya geometry}\label{sec_GVG}
In this subsection, we express the Vaidya metric in the thin shell
limit. Consider a $(d+1)$-dimensional Vaidya metric in the
Eddington-Finkelstein coordinates
\begin{equation}
  ds^2=\frac{R_{AdS}^2}{z^2}\left(-f(v,z)dv^2-2dvdz+\sum_{i=1}^{d-1}dx_i^2\right).
\end{equation}
When the thickness of the null shell vanishes, the factor
$f(v,z)$ can be expressed as
\begin{equation}
  f(v,z)=1-\theta(v)g(z),
\end{equation}
where $\theta(v)$ is the step function. It means that for
$v < 0$, $f(v,z)=1$ while for $v>0$, $f(v,z)=h(z)=1-g(z)$. 
In addition, the corresponding temperature and the entropy density of the final equilibrium state are given by
\begin{equation}
  T=\frac{\left|h^{\prime}\left(z_{h}\right)\right|}{4 \pi}, \qquad s_{\mathrm{eq}}=\frac{R_{AdS}^{d-1}}{z_{h}^{d-1}} \frac{1}{4 G_{N}}.
\end{equation}
 Here we
work with any spacetime dimension $d\geq2$ and only require the
function $g(z)$ satisfying the following properties:
\begin{itemize}
  \item $g(z)=1$ at the horizon $z=z_h$.
  \item $g(z)$ increases monotonically with $z$, for $z<z_h$.
  \item $g(z)\rightarrow \omega z^{d}$ for $z\rightarrow 0$, where $\omega$ is some constant.
\end{itemize}
Within this setup, the SAdS black hole with
\begin{equation}
  g(z)=mz^d,
\end{equation}
and the RN-AdS black hole with
\begin{equation}
  g(z)=mz^d-q^2z^{2d-2}, \quad d\geq3
\end{equation}
as well as other generic black holes subject to these
properties can be covered.
\subsection{Strip as the subregion on the boundary}
For a given $(d-1)$-dimensional strip $\mathcal{A}$ on the
boundary, it can be parameterized by the coordinates
$(x,y_{1},...,y_{d-2})$. We assume that it has a finite width
along $x$ direction such that $x\in [-l,l]$ and infinite length
$L\rightarrow \infty$ along the directions of $y_i$ such that
$y_{i}\in[-\frac{L}{2},\frac{L}{2}]$, where $i=1,...,d-2$. Then
the area of the strip $A_\mathcal{A}$ can be expressed as
\begin{equation}\label{eq_area}
  A_\mathcal{A} = \int_{-\frac{L}{2}}^{\frac{L}{2}}d\stackrel{\rightarrow}{y}\int_{-l}^{l}dx.
\end{equation}
Next, we define the corresponding HRT surface and figure out
some relations which are useful for computing the on-shell volume in
subsection \ref{sec_V}.

\subsection{Solutions for corresponding HRT surface}
For a given boundary strip $\mathcal{A}$, the corresponding HRT surface $\gamma_{\mathcal{A}}$ can be parameterized by $z(x)$ and $v(x)$ with the boundary conditions
\begin{equation}\label{eq_bc1}
v(\pm l)=t,\quad z(\pm l)=0,
\end{equation}
where $t$ is the time measured by inertial observers on the
boundary.

At the tip of the HRT surface,  we have
\begin{equation}\label{eq_bc2}
  v'(0)=z'(0)=0,\quad z(0)=z_t,\quad v(0)=v_t,
\end{equation}
where $v_t$ and $z_t$ label the tip of HRT surface
$\gamma_{\mathcal{A}}$ at boundary time $t$. Then the induced
metric on the HRT surface $\gamma_{\mathcal{A}}$ can be expressed
as
\begin{equation}
 ds^{2}=\frac{R_{AdS}^2}{z^{2}}\left[-f(v,z)v'^{2}-2z'v'+1\right]dx^{2}+\frac{R_{AdS}^2}{z^{2}}\sum_{i=1}^{d-2}dy_{i}^{2}.
\end{equation}
The area functional of the surface $\gamma_{\mathcal{A}}$
is given by
\begin{equation}\label{eq_At}
  A(t)=R_{AdS}^{d-1}L^{d-2}\int_{-l}^{l}\mathcal{L_{S}}\,dx,\qquad \mathcal{L_S}:=\frac{\sqrt{1-f(v,z)v'^{2}-2z'v'}}{z^{d-1}}.
\end{equation}
Treating this integral as the action, the corresponding
equations of motion (E.O.M) are
\begin{align}\label{eq_eom1}
  z^{2d-2}\mathcal{L_S}\partial _x\left(\frac{z'+fv'}{z^{2d-2}\mathcal{L_S}}\right)&=\frac{1}{2}\partial _v fv'^2,\\\label{eq_eom2}
  z^{2d-2}\mathcal{L_S}\partial _x\left(\frac{v'}{z^{2d-2}\mathcal{L_S}}\right)&=(d-1)z^{2d-3}\mathcal{L_S}^2+\frac{1}{2}\partial _z fv'^2.
\end{align}
The solutions to these EOM determine the configuration of the HRT surface $\gamma_{\mathcal{A}}$. Since the Lagrangian $\mathcal{L_S}$ does not depend on $x$ explicitly, we find a conserved quantity
\begin{equation}\label{eq_C}
  z^{d-1}\sqrt{1-f(v,z)v'^{2}-2z'v'}=C.
\end{equation}
In addition, when $f(v,z)$ does not depend on $v$ explicitly, we find anothor conserved quantity
\begin{equation}\label{eq_E}
  z'+f(z)v'=E.
\end{equation}
\begin{itemize}
  \item In the AdS region, from (\ref{eq_bc2}) and (\ref{eq_E}) we obtain
\begin{equation}\label{eq_AdSvz}
  E=0, \quad \frac{d\tilde{v}}{d\tilde{z}}=-1,
\end{equation}
where the symbol ``\;$\mathbb{\tilde{}}$\;'' represents the solutions of HRT surface. From (\ref{eq_bc1}), (\ref{eq_bc2}) and (\ref{eq_C}) we have
\begin{equation}\label{eq_adsz'}
C=z_{t}^{d-1}, \quad  z'=-\sqrt{\left(\frac{z_{t}}{z}\right)^{2d-2}-1}.
\end{equation}
Integrating the second expression of (\ref{eq_adsz'}), we
have
\begin{equation}\label{eq_AdSx}
  \tilde{x}(z)=\int_{z}^{z_t}dy\frac{1}{\sqrt{\left(\frac{z_t}{y}\right)^{2d-2}-1}}.
\end{equation}

\item Consider the matching conditions. The coordinates
between the AdS region and the black hole region should be
continuous. We further denote the location of the intersection
between HRT surface $\gamma_{\mathcal{A}}$ and the null shell
$v=0$ as $(z_c,0)$. Thus, in the AdS region we have
\begin{align}
  &z_c=z_t+v_t,\\
  z'_{-}=-v'_{-}&=-\sqrt{\left(\frac{z_t}{z_c}\right)^{2d-2}-1}.
\end{align}
Then, by integrating E.O.M (\ref{eq_eom1}) and (\ref{eq_eom2}), we
obtain the relations in black hole region, which are
\begin{align}
  &v'_+=v'_-,\\\label{eq_z'+}
  z'_+=&\left(1-\frac{1}{2}g(z_c)\right)z'_-.
\end{align}
Here the subscript $+ (-)$ refers to the derivatives on the black
hole (AdS) side. \item In black hole region, we obtain the
conserved quantities from the matching conditions, which are
\begin{equation}
  C=z_{t}^{d-1}, \quad E(z_c,z_t)=-\frac{1}{2}g(z_c)\sqrt{\left(\frac{z_{t}}{z_c}\right)^{2d-2}-1}.
\end{equation}
Substituting them into (\ref{eq_C}) and (\ref{eq_E}), we
have
\begin{equation}\label{eq_bhz'}
  z'^{2}=h(z)\left(\left(\frac{z_t}{z}\right)^{2d-2}-1\right)+E^2=:H(z),
\end{equation}
and
\begin{equation}
  v'=\frac{E-z'}{h(z)},
\end{equation}
which gives rise to the following relations in the black hole
region:
\begin{itemize}
  \item if $z'(x)\leq 0$ when $x>0$, then we have
\begin{align}\label{eq_bhvz1}
  \frac{d\tilde{v}}{d\tilde{z}}&=-\frac{1}{h(z)}\left(\frac{E}{\sqrt{H(z)}}+1\right),\\
  \label{eq_bhx}
  \tilde{x}(z)&=l-\int_{0}^{z}\frac{dy}{\sqrt{H(y)}}.
\end{align}
  \item if $z'(x)\geq 0$ when $x>0$, (\ref{eq_bhvz1}) should be modified as
  \begin{align}\label{eq_bhvz2}
    \frac{d\tilde{v}}{d\tilde{z}}&=\frac{1}{h(z)}\left(\frac{E}{\sqrt{H(z)}}-1\right),
  \end{align}
  and the new form of (\ref{eq_bhx}) depends on the integral.
\end{itemize}
\end{itemize}
In subsection \ref{sec_V}, we will see that both expressions of
$\frac{d\tilde{v}}{d\tilde{z}}$ and $\tilde{x}(z)$ play important
roles in deriving the on-shell volume. In the next subsection, we
tend to describe the configuration of critical HRT surfaces.
Taking $(3+1)$-dimensional SAdS case as an example, we show that
these relations are important when we investigate the intermediate
stage during the evolution of HSC as shown in section
\ref{sec_post}.

\begin{figure}
  \centering
  \subfigure[]{\label{fig_Hz1}
  \includegraphics[width=200pt]{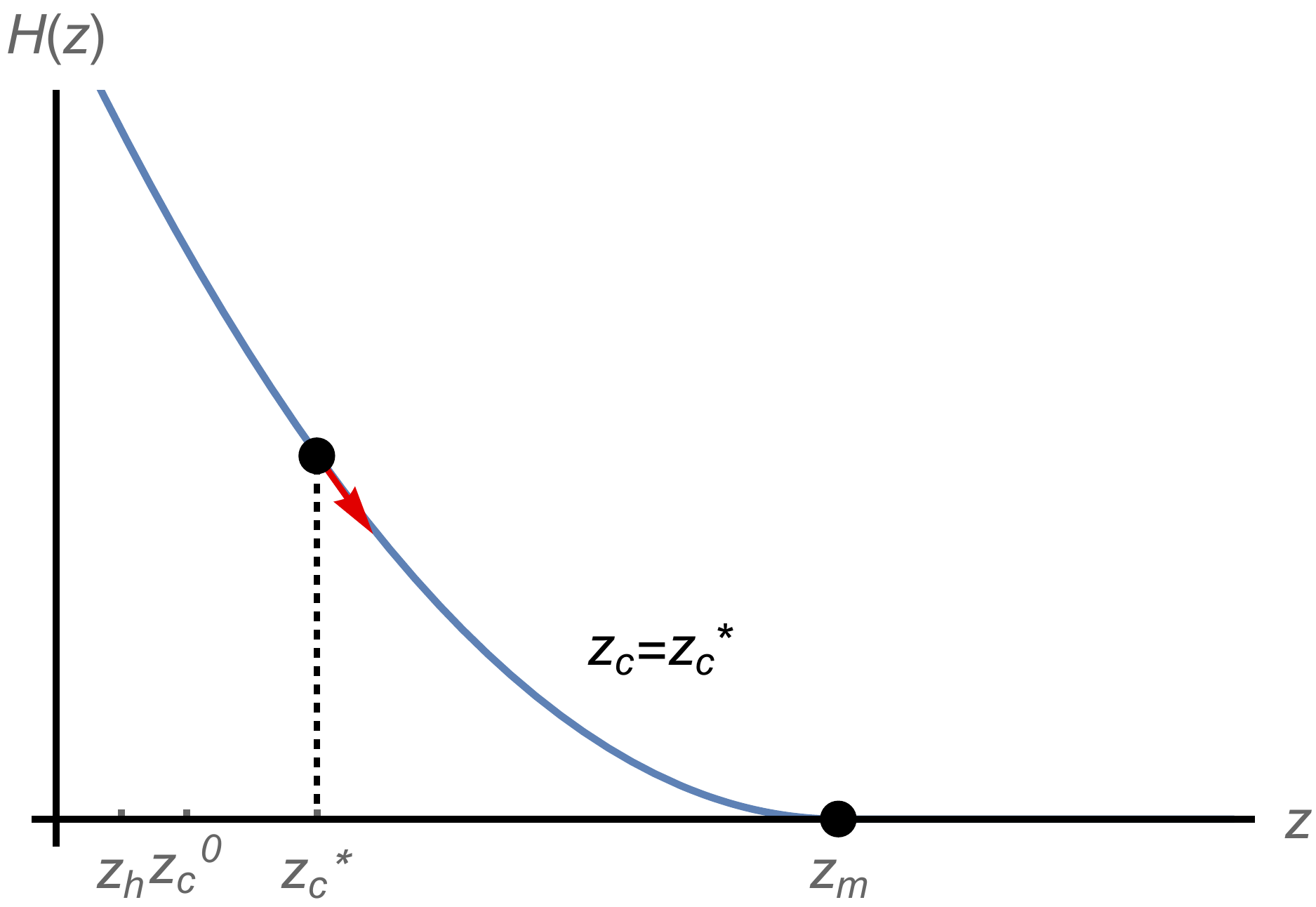}}
  \hspace{0pt}
  \subfigure[]{\label{fig_zx1}
  \includegraphics[width=200pt]{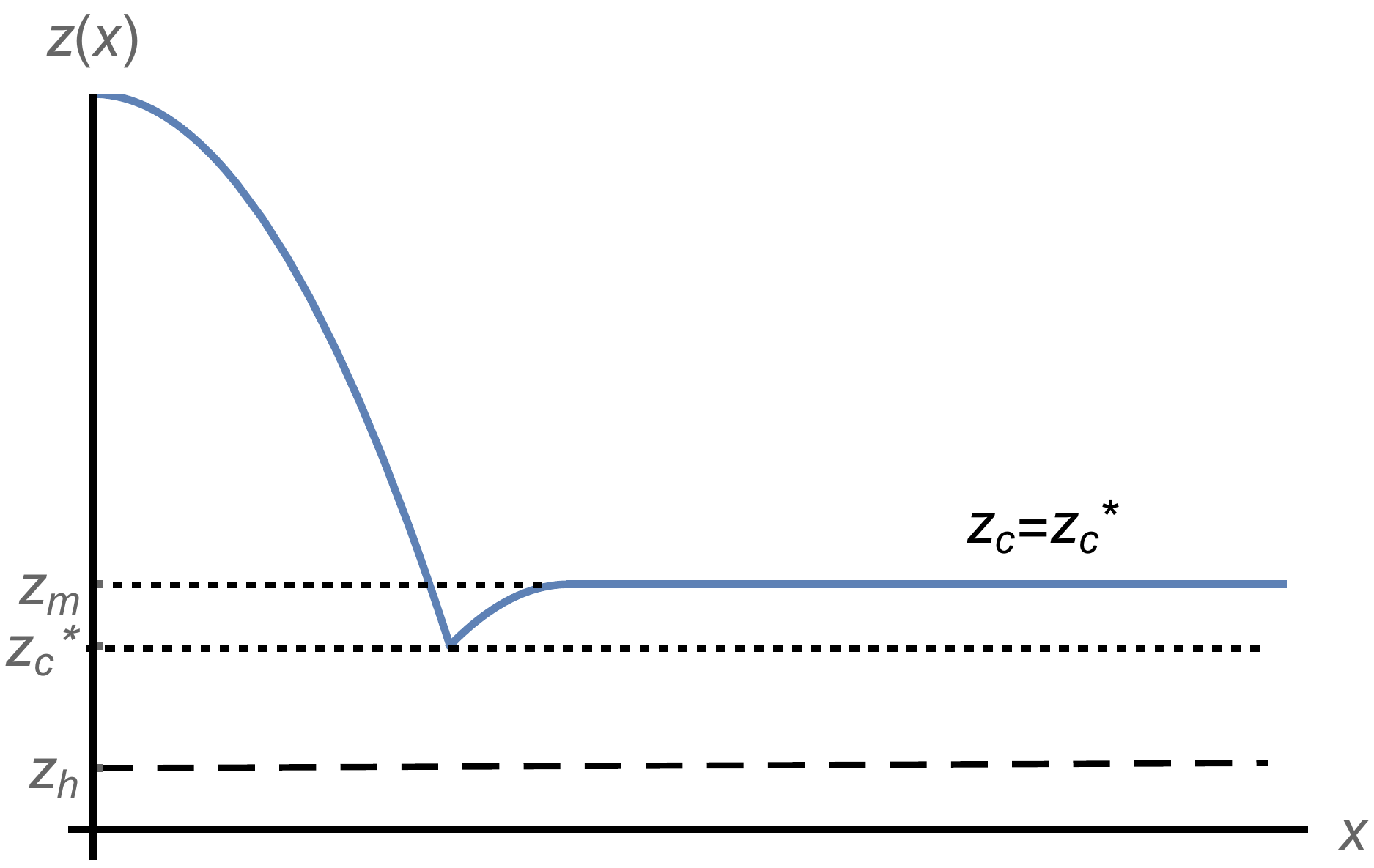}}

  \caption{For $z_c=z_c^*>z_c^0$ and $z_t\rightarrow \infty$, $H(z_m)=0$, and $H(z)$ decreases from $z=z_c^*$ to $z=z_m$ as shown in (a), which means $z(x)$ increases monotonically from $z_c$ and asymptotes to $z_m$. The corresponding configuration of the critical surface is shown in (b).}\label{fig_zc1}
\end{figure}

\subsection{Critical HRT surfaces}\label{sec_chs}
\begin{figure}
  \centering
  \subfigure[]{\label{fig_Hz2}
  \includegraphics[width=200pt]{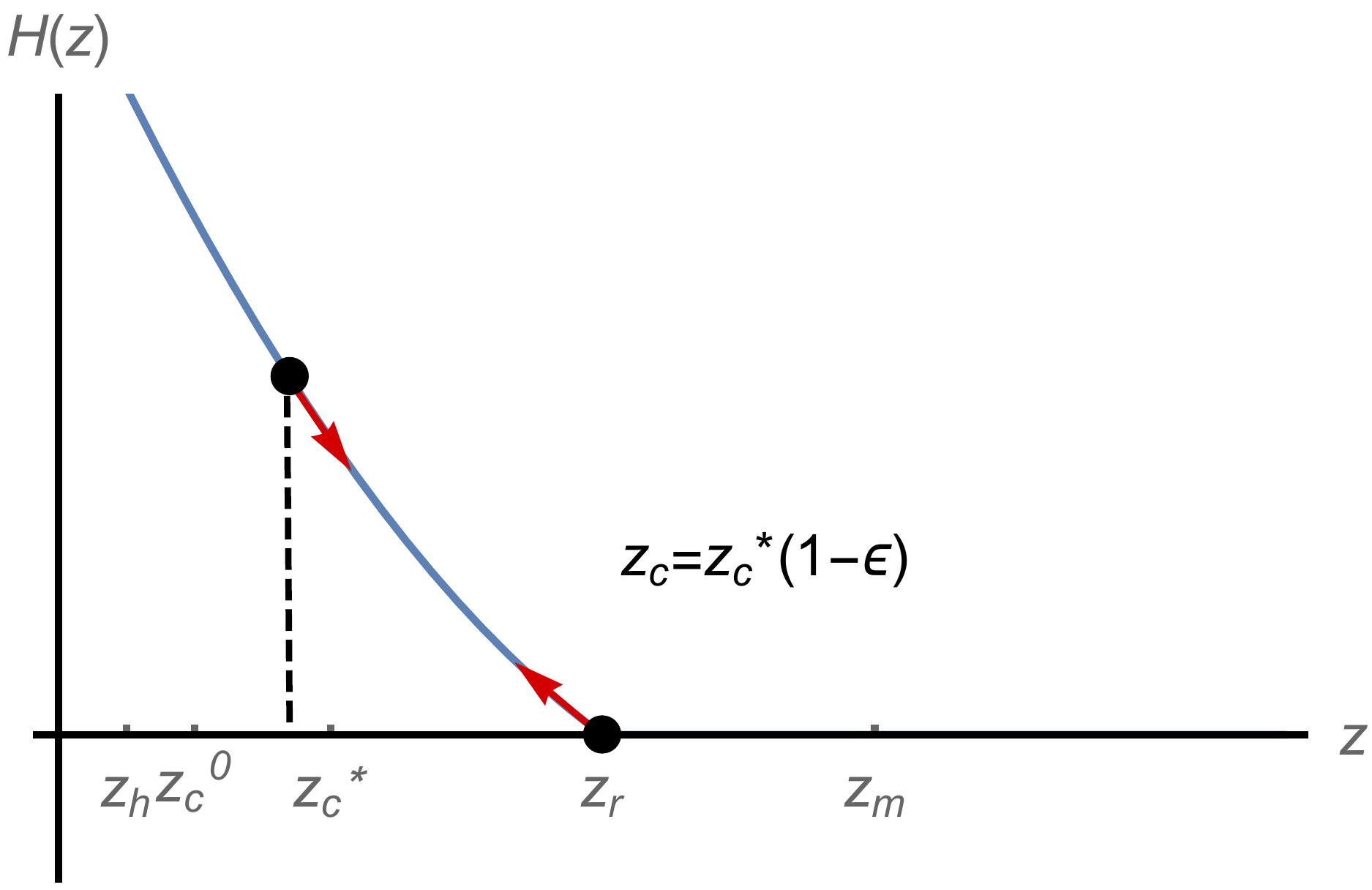}}
  \hspace{0pt}
  \subfigure[]{\label{fig_zx2}
  \includegraphics[width=200pt]{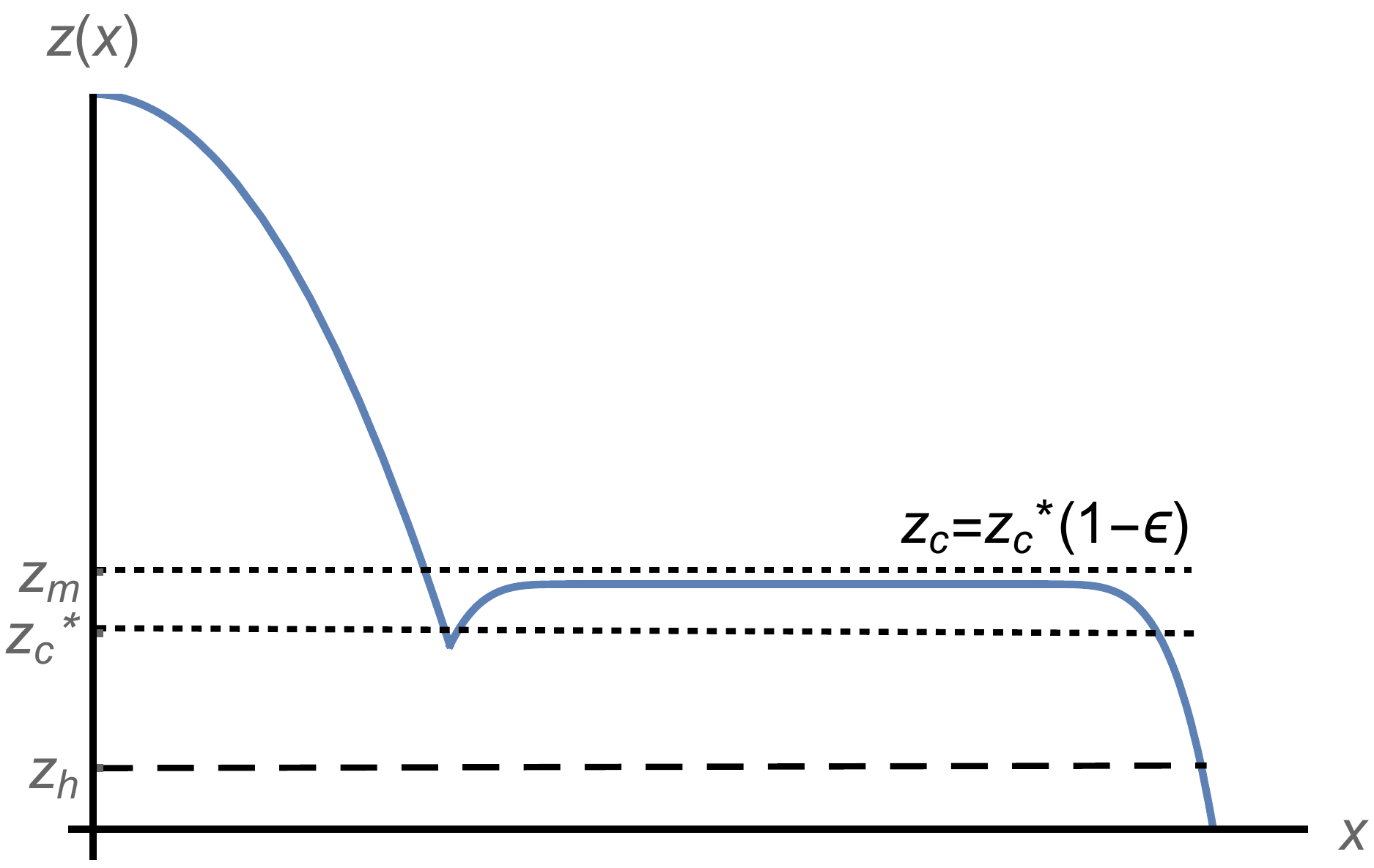}}

  \caption{For $z_c^0<z_c=z_c^*(1-\epsilon)$ where $0<\epsilon\ll 1$ and $z_t\rightarrow \infty$, $H(z_r)=0$, where $z_r$ is slightly less than $z_m$. $H(z)$ decreases from $z=z_c$ to $z=z_r$ as shown in (a), which means $z(x)$ increases from $z_c$ to a maximum $z_r$ and then decreases to zero. The corresponding configuration of the critical surface is shown in (b).}\label{fig_zc2}
\end{figure}

In \cite{Liu:2013qca}, it is shown that during the evolution of
HRT surface $\gamma_{\mathcal{A}}$, when $z_c$ goes to its
critical value called $z_c^*$, the HRT surface
$\gamma_{\mathcal{A}}$ approaches a critical configuration as
well. As shown in Fig.\ref{fig_zc1} and Fig.\ref{fig_zc2}, the
HRT surface $\gamma_{\mathcal{A}}$ surface reaches the boundary
only when $z_c<z_c^*$. In this work, we mainly focus on the
evolution of HRT surface under the condition of $z_c<z_c^*$.

Here we take SAdS with $h(z)=1-mz^3$ as the example, but the scheme involved in is also applicable to other cases, as discussed in \cite{Liu:2013qca}. Note that the first term in (\ref{eq_bhz'}) equals zero both at $z=z_h$ and $z=z_t$, and is negative in this interval. Therefore, there is a minimum of $H(z)$ between $z_h$ and $z_t$ which we denote $z=z_m$, i.e. $H'(z_m)=0$ and obtain an equation as
\begin{align}
  z_t^4=\frac{3m\,z_m^7}{4-m\,z_m^3}.
\end{align}
Now $z_c^*$ can be defined as
\begin{equation}
  H(z_m)\Big|_{z_c=z_c^*}=0.
\end{equation}
The solutions of above two equations are complicated in general, but in the limit of $z_t\rightarrow \infty$, the solutions reduce to
\begin{align}
  z_m=4^{1/3}m^{-1/3},\quad z_c^*=\frac{\sqrt{3}}{2^{1/3}}m^{-1/3},
\end{align}
where we have assumed that $z_t/z_m\gg1$ and $z_t/z_c^*\gg1$.
Note that in (\ref{eq_z'+}), if $g(z_c)>2$, then $z'_+>0$. Thus there exists a value $z_c^0$ such that $g(z_c^0)=2$.  Here we find that $z_c^0=2^{1/3}m^{-1/3}$. Then these quantities satisfy the  following relations
\begin{equation}\label{eq_zrela}
  z_h<z_c^0<z_c^*<z_m,
\end{equation}
where $z_h=m^{-1/3}$.

By setting $z_c=z_c^*$, we know $z_c^0<z_c=z_c^*$. Therefore, $z'_+$ is positive, indicating that $z$ increases with $x$ near the intersection. Next, we draw the configuration of this critical surface. As shown in Fig.\ref{fig_Hz1}, $z$ increases monotonically with $x$ from $z=z_c^*$ to $z=z_m$. In addition, Fig.\ref{fig_zx1} shows the configuration of the critical surface at $z_c=z_c^*$.

In the case of $z_c\rightarrow z_c^*$, we also have $z_c^0<z_c$ and $z'_+>0$. Firstly, note that
\begin{align}
  \frac{d}{dz_c}E^2=\frac{1}{2}m^2z_c(-3z_c^4+z_t^4).
\end{align}
In the limit of $z_t\gg z_c$, we have
\begin{equation}\label{eq_e^2}
  \frac{d}{dz_c}E^2>0.
\end{equation}
Combining (\ref{eq_bhz'}) with (\ref{eq_e^2}) we find that when
$z_c=z_c^*(1-\epsilon)$, where $\epsilon$ is small and positive,
the relation shows $H(z_m)\Big|_{z_c=z_c^*(1-\epsilon)}<0$. As
shown in Fig.\ref{fig_Hz2}, $z$ increases monotonically with $x$
from $z=z_c$ to $z=z_r$\footnote{Here we denote $z_r$ as a root of
$H(z)\Big|_{z_c=z_c^*(1-\epsilon)}=0$.} at first and then
decreases monotonically with $x$ from $z=z_r$ to $z=0$. In
Fig.\ref{fig_zx2} we show the configuration of the near critical
extremal surface in this case. Furthermore, we notice that the
closer $z_c$ gets to $z_c^*$, the longer the surface hangs along
the $x$ direction.

After obtaining the configuration of the HRT surface near the critical position, we can generalize the above discussion to the HSC and then analyze its growth behavior.

\subsection{The on-shell volume with translational symmetry}\label{sec_V}

The $d$-dimensional bulk extremal surface $\Gamma_{\mathcal{A}}$, which is bounded by the boundary strip $\mathcal{A}$ and the corresponding HRT surface $\gamma_{\mathcal{A}}$, can be parameterized by coordinates
 $(z,x,y_{1},...,y_{d-2})$. Under this parameterization, the volume functional $V(t)$ of the codimension-one extremal surface $\Gamma_{\mathcal{A}}$ can be expressed as
\begin{equation}\label{eq_invoz}
  V(t)=2\int_{0}^{z_t}dz\int_{-\frac{L}{2}}^{\frac{L}{2}}d\stackrel{\rightarrow}{y}\int_{0}^{\tilde{x}(z)}dx\mathcal{L}(z,v(z,x,\stackrel{\rightarrow}{y})),
\end{equation}
where
$\tilde{x}(z)$
is the solution
 of the corresponding HRT surface.

Taking the translational symmetry into account, we rewrite (\ref{eq_invoz}) into a simple form
\begin{align}
  V(t)=2R_{AdS}^d\int_{0}^{z_t}dz\int_{-\frac{L}{2}}^{\frac{L}{2}}d\stackrel{\rightarrow}{y}\int_{0}^{\tilde{x}(z)}dx\mathcal{L}(z,v(z)).
\end{align}
Then the induced metric on the the surface $\Gamma_{\mathcal{A}}$ can be expressed as
\begin{equation}
  ds^2=\frac{R_{AdS}^2}{z^2}\left[-\left(f(v,z)\frac{dv}{dz}+2\right)\frac{dv}{dz}dz^2+dx^2+d\stackrel{\rightarrow}{y}^2\right].
\end{equation}
We can read the volume functional from the above induced metric, that is
\begin{align}\label{eq_Voz}
  &V(t)=2R_{AdS}^dL^{d-2}\int_{0}^{z_t}dz\int_{0}^{\tilde{x}(z)}dx\sqrt{-f(v(z),z)\left(\frac{d v}{d z}\right)^{2}-2\frac{d v}{d z}} z^{-d},\\
  &\mathcal{L}(z,v(z)):=\sqrt{-f(v(z),z)\left(\frac{d v}{d z}\right)^{2}-2\frac{d v}{d z}} z^{-d}.\nonumber
\end{align}

Substituting (\ref{eq_AdSvz}), (\ref{eq_AdSx}) and (\ref{eq_bhvz1}), (\ref{eq_bhx}) into (\ref{eq_Voz}), we find it reduces to the  expression of the on-shell volume.

\section{Three characteristic stages}\label{sec_GCS}
So far, we have obtained the explicit form of the on-shell volume for the extremal surface with translational symmetry. In this section, we will describe the evolution behavior of HSC during the quench by dividing the whole process into three distinct stages. At each stage, we will analytically compute the HSC with a focus on its dynamical behavior.
\subsection{Early-time growth stage}
In this subsection, we consider the evolution of the on-shell volume $V(t)$ in the early time, which implies
\begin{equation}
  t\ll z_h= \frac{R_{AdS}}{(4 G_N)^{\frac{1}{d-1}}} s_{\mathrm{eq}}^{\frac{1}{1-d}}.
\end{equation}
In this period, the null shell and the crossing point $(z_c,0)$ are near the boundary. As a result we are allowed to expand the expression of $V(t)$  with small $t$ and then  investigate its growth behavior.

To the leading order of small $t$, the change of the on-shell volume (\ref{eq_Voz}) can be expressed as

\begin{equation}\label{} \frac{\Delta V(t)}{2R_{AdS}^dL^{d-2}}=A_{I}+A_{II}+A_{III}+A_{IV}\cdots,
\end{equation}
with
\begin{align}\label{eq_a1i}
  A_{I}=&\left(\int_{0}^{z_t}dz\int_{0}^{\tilde{x}(z)}dx\frac{\partial \mathcal{L}}{\partial f}\right)\Bigg|_{t=0}\delta f,\\\label{eq_a2i}
  A_{II}=&\frac{\partial}{\partial z_t}\left(\int_{0}^{z_t}dz\int_{0}^{\tilde{x}(z)}dx\mathcal{L}\right)\Bigg|_{t=0}\delta z_t,\\\label{eq_a3i}
  A_{III}=&\int_{0}^{z_0}dz\,\tilde{x}_{0}(z)\frac{\partial \mathcal{L}}{\partial v'}\Bigg|_{t=0}\delta v',\\
  A_{IV}=&\int_{0}^{z_0}dz\,\mathcal{L}\Bigg|_{t=0}\delta \tilde{x}(z),
\end{align}
where $z_0$ denotes the value of $z_t$ at $t=0$, and $\tilde{x}_{0}(z)$ denotes $\tilde{x}(z)$ at $t=0$, while $v'$ denotes the derivative of $v(z)$ with respect to $z$ (which is different from the case in holographic entanglement entropy).

Then we substitute the Lagrangian $\mathcal{L}$ (\ref{eq_Voz})
into (\ref{eq_a1i}), (\ref{eq_a2i}) and (\ref{eq_a3i})
respectively. Since $\delta f\neq 0$ only when $z\in[0,z_c]$,
(\ref{eq_a1i}) is reduced to
\begin{equation}\label{eq_a1}
  A_{I}=\frac{\omega}{2}\int_{0}^{z_c}dz\,\tilde{x}_{0}(z)+\cdots = \frac{1}{2}\omega lt\cdots.
\end{equation}
In the last step, it is not difficult to show that for $t\rightarrow 0$, we have $z_c\rightarrow t$ and $\tilde{x}_{0}(z)\rightarrow l$.

Next we claim that the contributions from the remaining terms,
namely $A_{II}$, $A_{III}$  and $A_{IV}$, are vanishing. Firstly,
\begin{equation}\label{eq_a21}
  A_{II}=\tilde{x}_0(z_0)\mathcal{L}\Bigg|_{z=z_0}\delta z_t.
\end{equation}
This term is vanishing since $\tilde{x}(z)=0$ at the tip $z=z_t$. While for the term $A_{III}$, we find that
\begin{equation}\label{eq_a31}
  \frac{\partial \mathcal{L}}{\partial v'}\Bigg|_{t=0}= \frac{z^{-d}}{2}\frac{-2fv'-2}{\sqrt{-fv'^{2}-2v'}}\Bigg|_{t=0}=0.
\end{equation}
Since  $f\big|_{t=0}=1$ and $v'\big|_{t=0}=-1$, the term $A_{III}$ vanishes as well.

Now, for the last term $A_{IV}$,
\begin{align}\label{eq_A4}
  A_{IV}=&\int_{t}^{z_0}\frac{dz}{z^d}\left(z_0\int_{z/z_0}^{1}\frac{dy}{\sqrt{y^{2-2d}-1}}-z_t\int_{z/z_t}^{1}\frac{dy}{\sqrt{y^{2-2d}-1}}\right)\nonumber\\
  &+\int_{0}^{t}\frac{dz}{z^d}\,\int_{0}^{z}dy \left(\frac{1}{\sqrt{H(y)}}-\frac{1}{\sqrt{\left(\frac{z_{0}}{y}\right)^{2 d-2}-1}}\right).
\end{align}

We would like to define
\begin{align}
  f_1(z_t,z)&:=z_t\int_{z/z_t}^{1}\frac{dy}{\sqrt{y^{2-d}-1}},\\
  F_2(t)&:=\int_{0}^{t}\frac{dz}{z^d}\,\int_{0}^{z}dy \left(\frac{1}{\sqrt{H(y)}}-\frac{1}{\sqrt{\left(\frac{z_{0}}{y}\right)^{2 d-2}-1}}\right),
\end{align}
then it is convenient to expand (\ref{eq_A4}) as
\begin{align}\label{eq_A41}
  A_{IV}=&\int_{0}^{z_0}\frac{dz}{z^d}\frac{\partial f_1}{\partial z_t}\Bigg|_{t=0}\delta
  z_t+F'_2\Bigg|_{t=0}t+\cdots.
\end{align}

For the first term in (\ref{eq_A41}), we notice that
$\int_{0}^{z_0}\frac{dz}{z^d}\frac{\partial f_1}{\partial z_t}|_{t=0}$ is
independent of $t$. Thus we only consider the variation of
$z_t$, i.e. $\delta z_t=z_{0}-z_{t}$. Since the half-width $l(t)$ of
boundary strip is conserved during time evolution, namely
$l(0)=l(t)$, from this equation we have
\begin{equation}
\int_{0}^{z_0}\frac{dz}{\sqrt{\left(\frac{z_0}{z}\right)^{2d-2}-1}}=\int_{t}^{z_t}\frac{dz}{\sqrt{\left(\frac{z_t}{z}\right)^{2d-2}-1}}+\int_{0}^{t}\frac{dz}{\sqrt{H(z)}}.
\end{equation}
Then for $t\rightarrow 0$, the variation of $z_t$ can be expressed
as
\begin{equation}
  \delta z_t=\frac{\Gamma(\frac{1}{2(d-1)})}{\sqrt{\pi}\Gamma(\frac{d}{2(d-1)})}\left[f_2'(t)\Bigg|_{t=0}t+\frac{f_2''(t)}{2}\Bigg|_{t=0}t^2+\cdots\right],
\end{equation}
where
$$f_2(t)=\int_{0}^{t} d z\left[\frac{1}{\sqrt{H(z)}}-\frac{1}{\sqrt{\left(\frac{z_{t}}{z}\right)^{2 d-2}-1}}\right].$$
In the above equations, we find that $f_2'(t)\big|_{t=0}$ equals
zero. Therefore, the variation of $\delta z_t$ is at least of
order $\mathcal{O}(t^2)$.

As for the second term in (\ref{eq_A41}), we also find that
$F_2'(t)\big|_{t=0}$ equals zero. Therefore,
\begin{equation}
  A_{IV}\approx \mathcal{O}(t^2).
\end{equation}

Based on the discussion above, we finally obtain the
leading-order behavior of $\Delta V(t)$ as
\begin{equation}\label{eq_Vtf}
  \Delta V(t)=2R_{AdS}^dL^{d-2}l\frac{1}{2}\omega t+\cdots = A_{\mathcal{A}}R_{AdS}^d\frac{1}{2}\omega t+\cdots.
\end{equation}
The result indicates that at the very early time when the null
shell starts to fall down, the HSC grows linearly with time no
matter how small the size of the subregion is, but the rate of growth is sensitive to the size of the system as well as the asymptotic form of the metric. Since the UV modes
of a mixed state $\mathcal{A}$ become thermalized first in
this setting (see also \cite{Balasubramanian:2010ce,
Balasubramanian:2011ur}), at the early time, the UV parts of
different states $\mathcal{A}$ change with the same rate and
contribute equally to the growth of complexity. This result
also reveals that during the thermalization process, the
increase of complexity is sensitive to the high-momentum modes of different quantum ensembles.

In particular, when the size of the region $\mathcal{A}$ goes
to infinity, the results of CV conjecture and its version of the
subregion can be compared directly as follows:

\begin{itemize}
  \item In the SAdS case, if we take $l\rightarrow \infty$, then we have $\omega=z_h^{-d}$ and the physical total mass
  \begin{equation}\label{eq_Mm}
    M=\frac{(d-1)A_{\mathcal{A}}R_{AdS}^{d-1}}{16\pi G_Nz_h^d}.
  \end{equation}
  Therefore, in the early time, the growth rate of holographic complexity is
  \begin{equation}
    \frac{dC}{dt}=\frac{1}{R G_N}\frac{d}{dt}V(t)=\frac{R_{AdS}}{R}\frac{8\pi M}{d-1}.
  \end{equation}
  If one sets the arbitrary length scale as $R=R_{AdS}$, then the result reduces to
  \begin{equation}
    \frac{dC}{dt}=\frac{8\pi M}{d-1},
  \end{equation}
  which is consistent with the result in \cite{Chapman:2018dem}.

  In addition, if one sets the length scale as mentioned in \cite{Kim:2017qrq}:
  \begin{equation}\label{eq_RR}
    \frac{R_{AdS}}{R}=\frac{d-1}{4\pi^2},
  \end{equation}
where we have set $\hbar=1$. Then the result reduces to
\begin{equation}
  \frac{dC}{dt}=\frac{2M}{\pi},
\end{equation}
which is satisfied with the Lloyd bound at early time.
\end{itemize}
This is also in agreement with the results in the previous works
\cite{Chen:2018mcc,Chapman:2018dem}. Further, we find that whether the
Lloyd bound is violated is sensitive to the choice of the length
scale $R$.

\subsection{Intermediate growth stage}\label{sec_post}
In this section, we investigate the growth behavior of HSC while
the configuration of $\Gamma_\mathcal{A}$ is near the critical
extremal surface, which can be quantitatively considered as the
limit of $l\gg t\gg z_h= \frac{R_{AdS}}{(4 G_N)^{\frac{1}{d-1}}} s_{\mathrm{eq}}^{\frac{1}{1-d}}$ and $z_c=z_{c}^{*}(1-\epsilon)$, where
$\epsilon\ll 1$. We further assume $z_t$ is very large and subject
to the following relations
\begin{align}
  \frac{z_c^*}{z_t},\;\frac{z_m}{z_t},\;\frac{z_c^*}{|\log\epsilon|}\;\ll 1.
\end{align}
We adopt the strategy presented in \cite{Liu:2013qca}: We
firstly expand time $t$ and the on-shell volume $V$ in
terms of $1/z_t$ and $\epsilon$; then from the leading orders
of expansion $t(z_t,\epsilon)$ and $V(z_t,\epsilon)$, we find
$V=V(t)$ to obtain the evolution behavior of HSC.

Based on (\ref{eq_bhvz1}), (\ref{eq_bhvz2}) and the
discussion in Sec.\ref{sec_chs}, we can express time $t$ and half-width
$l$ as
\begin{align}\label{eq_ct}
  t&=\int_{z_r}^{z_c}\frac{1}{h(z)}\left(\frac{E}{-\sqrt{H(z)}}+1\right)dz+\int_{0}^{z_r}\frac{1}{h(z)}\left(\frac{E}{\sqrt{H(z)}}+1\right)dz,\\\label{eq_cl}
  l&=\int_{z_c}^{z_t}\frac{dz}{\sqrt{\left(\frac{z_t}{z}\right)^{2d-2}-1}}+\left(\int_{z_c}^{z_r}+\int_{0}^{z_r}\right)\frac{dz}{\sqrt{H(z)}}.
\end{align}
Recall that in this regime we have
\begin{equation}
  0<z_h<z_c<z_r,
\end{equation}
and $z_r$ is a root of $H(z)\Big|_{z_c=z_c^*(1-\epsilon)}=0$.
Therefore, two terms become divergent in (\ref{eq_ct}),
which are
\begin{equation}\label{eq_izh}
  \frac{1}{h(z)}\Bigg|_{z\rightarrow z_h}\rightarrow \infty, \qquad \quad
  \frac{1}{H(z)}\Bigg|_{z\rightarrow z_r}\rightarrow \infty.
\end{equation}

Firstly, the divergence caused by the first term of (\ref{eq_izh})
at $z=z_h$ can be subtracted by taking the cut-off carefully.
Therefore, we focus only on the divergence caused by the second
term of (\ref{eq_izh}) at $z=z_r$.  We expand $H(z)$ at
$z_c=z_c^*(1-\epsilon)$ and $z=z_m$, then we get
\begin{equation}
  H(z)\Bigg|_{\substack{z_c\rightarrow z_c^*\\z\rightarrow z_m}}=C_1(z-z_m)^2+C_2\epsilon+\cdots,
\end{equation}
where $C_1=\frac{1}{2}H''(z_m)$ and
$C_2=-\frac{dE^2}{dz_c}\Big|_{z_c^*}z_c^*$. As a result, the
integrand in (\ref{eq_ct}) can be expanded  at $z=z_m$ and
$z_c=z_c^*$, and the leading terms can be extracted as
\begin{equation}
  t=\frac{-E}{h(z_m)\sqrt{C_1}}\log\epsilon+ \cdots.
\end{equation}
In parallel, for the half-width $l$, we expand the integrand in
(\ref{eq_cl})  with $1/z_t\rightarrow 0$, $z=z_m$ and $z_c=z_c^*$,
leading to
\begin{equation}\label{eq_cl1}
  l=c_dz_t-\frac{1}{\sqrt{C_1}}\log\epsilon+\cdots,
\end{equation}
where we set $c_d:=\frac{\sqrt{\pi}\Gamma\left(\frac{d}{2(d-1)}\right)}{\Gamma\left(\frac{1}{2(d-1)}\right)}$. Then the time dependence of  $z_t$  can be derived  from (\ref{eq_cl1}) as
\begin{equation}\label{eq_chzt}
  z_t=\frac{1}{c_d}\left(l-\frac{h(z_m)}{E}t\right)+\cdots.
\end{equation}

Now we turn to the on-shell volume $V(t)$. For convenience we define a normalized volume $\tilde{V}(t)$ as
\begin{equation}\label{eq_nv}
  \tilde{V}(t)=\frac{V(t)}{2R_{AdS}^dL^{d-2}}=\tilde{V}_{AdS}+\tilde{V}_{BH}.
\end{equation}
The first term $\tilde{V}_{AdS}$ is
\begin{align}
  \tilde{V}_{AdS}&=L_dz_c^{*\,1-d}\left(l-\frac{h(z_m)}{E}t\right)+\cdots,\\
  L_d:&=\frac{\sqrt{\pi}\Gamma[(3+\frac{1}{d-1})/2]}{d(d-1)\,c_d\,\Gamma[1+\frac{1}{2(d-1)}]}.
\end{align}
It indicates that the volume in AdS part decreases linearly with $t$. Intuitively this is reasonable since the extremal surface $\Gamma_{\mathcal{A}}$ is moving towards the horizon region of the black hole during the evolution.

Now we turn to consider the second term $\tilde{V}_{BH}$ in (\ref{eq_nv}), which is
\begin{equation}\label{eq_bhnv}
  \tilde{V}_{BH}=\int_0^{z_c}\frac{dz}{z^d}\sqrt{-f\left(\frac{dv}{dz}\right)^2-2\frac{dv}{dz}}\tilde{x}(z).
\end{equation}
According to (\ref{eq_bhvz1}), (\ref{eq_bhvz2}) and the discussion
in Sec.\ref{sec_chs}, we rewrite (\ref{eq_bhnv}) into
\begin{equation}
  \tilde{V}_{BH}
  =\int_0^{z_r}\frac{dz}{z^d}\tilde{x}(z)\sqrt{-\frac{1}{h(z)}\frac{E^2}{H(z)}+\frac{1}{h(z)}} + \int_{z_c}^{z_r}\frac{dz}{z^d}\tilde{x}(z)\sqrt{-\frac{1}{h(z)}\frac{E^2}{H(z)}+\frac{1}{h(z)}}.
\end{equation}
Note that the divergence at $z=0$ is cancelled by subtracting the
vaccum value $\tilde{V}_{vac}$ and similar to the above
discussion, the integrand is regular at $z=z_h$. Therefore, we
should only consider the divergence at $z=z_r$. For this purpose
we expand the subtracted volume $\Delta\tilde{V}(t)$ at $z=z_m$
and $z_c=z_c^*$, which is
\begin{equation}\label{eq_cdv}
\Delta\tilde{V}(t)=\tilde{V}(t)-\tilde{V}_{vac}=\left(\frac{\tilde{x}(z_m)}{z_m^d}\frac{z_t^{d-1}}{z_m^{d-1}}-L_dz_c^{*\,1-d}\right)\frac{h(z_m)}{E}\,t+\cdots.
\end{equation}

The expression $\tilde{x}(z)$ at $z_m$ can be expressed as
\begin{equation}
  \tilde{x}(z_m)=l-\int_0^{z_m}\frac{d z}{\sqrt{H(z)}}=l-\frac{h(z_m)}{2E}\,t+\cdots.
\end{equation}
Therefore,  (\ref{eq_cdv})  can be rewritten as
\begin{equation}\label{eq_chV}
  \Delta V(t)=A_\mathcal{A}R_{AdS}^d\,\eta \,t+\cdots,\qquad\quad
  \eta =\left[\frac{l^{d-1}}{c_d^{d-1}z_m^{2d-1}}-\frac{L_d}{l\,z_c^{*\,d-1}}\right]\frac{h(z_m)}{E}.
\end{equation}

The result shows that at the intermediate stage, the HSC grows
linearly with time as well. But the situation is very
different from the one at the early stage. Firstly, the expression
is valid only when the size of the subregion $\mathcal{A}$ goes to
infinity. Secondly, the growth rate $\eta$ is sensitive to the
dimension of the system. Last but not least, different types of
quench give different rates of growth even if the UV modes of the
systems contribute uniformly. For instance, thermal quench or
electromagnetic quench will usually give different rates of growth
even if the metric of the background in both cases is
asymptotic to the AdS spacetime.

Next, we take the SAdS black hole as the example  to derive  the
growth rate explicitly and compare the result with the one under
CV conjecture. In this case the function in the metric is given as
\begin{equation}
  h(z)=1-\left(\frac{z}{z_h}\right)^d.
\end{equation}
Following the discussion in Sec.\ref{sec_chs},  we find in the
limit of $z_t\rightarrow \infty$, $z_m$ and $z_c^*$ can be
expressed as
\begin{equation}\label{eq_zz3}
  z_m=\left(\frac{2d-2}{d-2}\right)^{1/d}z_h, \qquad z_c^*=\frac{\sqrt{d(d-2)}}{d-1}\left(\frac{2d-2}{d-2}\right)^{1/d}z_h,
\end{equation}
with $d\geq 3$. When $d=2$, we have
\begin{equation}\label{eq_zz2}
  z_m=\sqrt{z_t z_h}, \qquad\qquad z_c^*=2z_h.
\end{equation}
Therefore, we discuss the growth rate in two different cases:
\begin{itemize}
  \item When $d\geq 3$, $z_m$ and $z_c^*$ are finite.
Note that in this case, $l$ goes to infinity while other coefficients remain finite. Therefore, we can rewrite $\Delta\tilde{V}(t)$ as
\begin{equation}
  \Delta\tilde{V}(t)=\frac{2l}{z_m^{2d-1}}\frac{ z_m^d-z_h^d}{z_c}\,t+\cdots
\end{equation}
and the corresponding subtracted volume $\Delta V(t)$ as
\begin{equation}\label{eq_nchV}
  \Delta V(t)=A_{\mathcal{A}}R_{AdS}^d\frac{2}{z_m^{2d-1}}\frac{ z_m^d-z_h^d}{z_c}\,t+\cdots.
\end{equation}
Substituting (\ref{eq_zz3}) into the above equation, we obtain
\begin{equation}
  \Delta V(t)=A_{\mathcal{A}}R_{AdS}^d\frac{\sqrt{d(d-2)}}{2(d-1)}z_h^{-d}\,t+\cdots.
\end{equation}

By virtue of (\ref{eq_Mm}) and (\ref{eq_RR}), the above expression reduces to
\begin{align}
  \frac{dC}{dt}=&\frac{1}{R G_N}\frac{d}{dt}V(t)
  =\frac{\sqrt{d(d-2)}}{d-1}\frac{R_{AdS}}{R}\frac{8\pi M}{d-1}.
\end{align}
In the same way, if we choose $R=R_{AdS}$, then the result becomes
\begin{equation}\label{eq_ltg1}
  \frac{dC}{dt}=\frac{\sqrt{d(d-2)}}{d-1}\frac{8\pi M}{d-1}.
\end{equation}
On the other hand, if we choose
\begin{equation}
  \frac{R_{AdS}}{R}=\frac{d-1}{4\pi^2},
\end{equation}
then the result becomes
\begin{equation}\label{eq_ltg2}
  \frac{dC}{dt}=\frac{\sqrt{d(d-2)}}{d-1}\frac{2M}{\pi}.
\end{equation}
We remark that the Lloyd bound is always satisfied since $$\frac{\sqrt{d(d-2)}}{d-1}<1$$ for $d\geq 3$, and the bound is saturated only when the dimension $d\rightarrow\infty$.

\item When $d=2$, $z_m$ goes to infinity but $z_c^*$ keeps finite.

First of all, though $z_m$ goes to infinity in this case, the condition
\begin{equation}
  \frac{z_m}{z_t}=\sqrt{\frac{z_h}{z_t}}\ll 1
\end{equation}
still holds. Thus, the discussion in Sec.\ref{sec_post} holds as
well. Next, we turn to analyze the growth rate in this case.
Substituting (\ref{eq_zz2}) and (\ref{eq_chzt}) into
(\ref{eq_chV}), we find the coefficients reduce to
\begin{equation}
  \frac{h(z_m)}{E}\approx\frac{z_m^2}{z_c^{*}z_t/2}=1.
\end{equation}
Then (\ref{eq_chV}) has a similar form as (\ref{eq_nchV})
\begin{equation}
  \Delta V(t)=A_{\mathcal{A}}R_{AdS}^2\frac{z_h^{-3/2}}{\sqrt{l}}t+\cdots.
\end{equation}
The corresponding growth rate of HSC is
\begin{align}
  \frac{dC}{dt}=&\frac{1}{RG_N}\frac{d}{dt}V(t)
  =\frac{R_{AdS}}{R}\sqrt{\frac{4z_h}{l}}8\pi M+\cdots.
\end{align}

Note that when $l\rightarrow \infty$, the linear order of the
growth rate is actually going to zero near the critical
configuration. This phenomenon indicates that in the
$(2+1)$-dimensional SAdS case, the change rate of HSC is dominated
by the higher order terms of $t$ at the intermediate stage, which can be numerically checked as shown in Fig. \ref{fig_evot}.

\begin{figure}
  \centering
  \subfigure[]{\label{fig_ztt}
  \includegraphics[width=225pt]{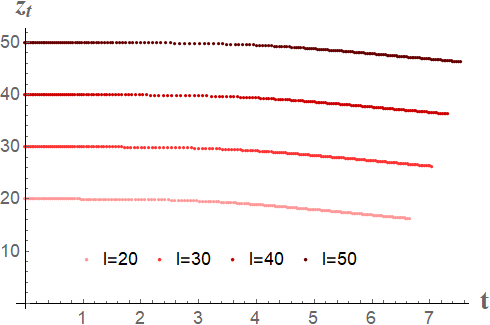}}
  \hspace{0pt}
  \subfigure[]{\label{fig_ct}
  \includegraphics[width=225pt]{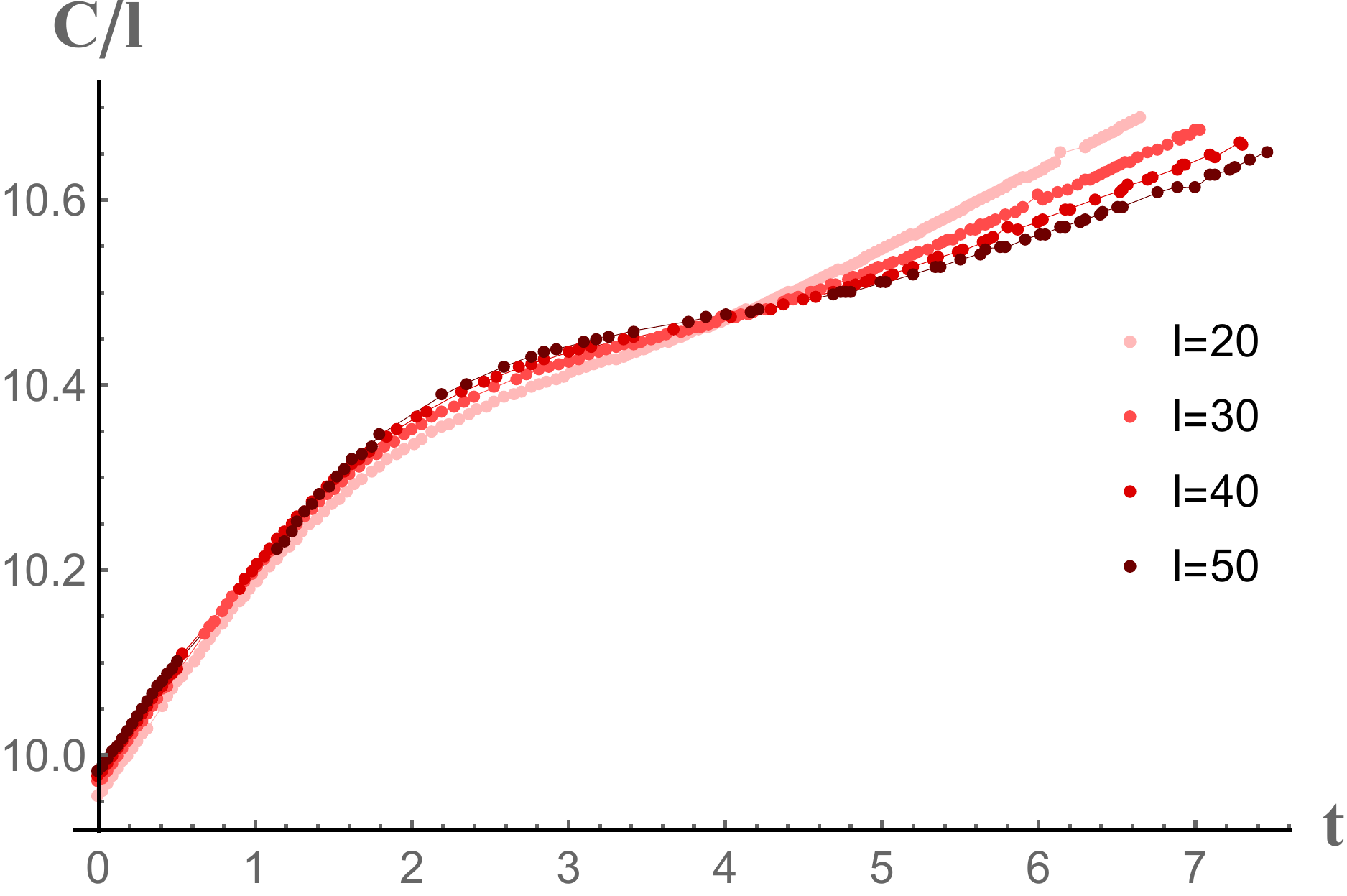}}

  \caption{Different red lines represent the subsystems with different half-width $l$. The evolution of the tip $z_t$ of the extremal surface is shown in \ref{fig_ztt} and the evolution of the HSC at the intermediate stage (which is near $t=4$) is shown in \ref{fig_ct}. For $l\approx z_t\gg t$, the rate of growth approaches zero as the half-width $l$ goes to infinity. For large $l$, the numerical simulations are credible at the intermediate stage, but become out of control at the late-time stage.}\label{fig_evot}
\end{figure}

\end{itemize}

We notice that the above results are very different from those
of the CV conjecture \cite{Chapman:2018dem}. In the CV conjecture, the ``tip'' $z_t$ of the extremal surface is always anchored at infinity. While in the subregion CV conjecture, the tip $z_t$ of the extremal surface $\Gamma_{\mathcal{A}}$ decreases with time $t$, according to (\ref{eq_chzt}). In general, different configurations of extremal surfaces lead to different growing behaviors. As we can see in the next section, the difference in the configurations of extremal surfaces will become significant at the late time and so as the rate of growth.

\subsection{Late-time stage of the evolution}

\subsubsection{Discontinuous transition at the late time}
In the cases of SAdS with $d\geq 3$, the discussion in
Sec.\ref{sec_post} is naturally generalized to the late-time
stage. For a region $\mathcal{A}$ with the half-width $l$, given
the linear growth (\ref{eq_nchV}), the equilibrium time of the
evolution is \footnote{We directly apply the result firstly
obtained in
\cite{Liu:2013qca} here, since the evolutions of $\gamma_{\mathcal{A}}$ and
$\Gamma_{\mathcal{A}}$ share a common equilibrium
time\cite{Chen:2018mcc}.}
\begin{equation}\label{eq_ts0}
  t_s=\frac{4 G_N l}{R_{AdS}^{d-1} z_m^{1-d}\sqrt{-h(z_m)}}s_{eq}.
\end{equation}
From (\ref{eq_chzt}) and (\ref{eq_ts0}), the tip $z_t$ at the
equilibrium time $t_s$ reduces to
$$z_t=\frac{l}{c_d}\left[1-\left(\frac{z_m^2}{z_tz_h}\right)^{d-1}\right]+\cdots.$$
For large $z_t$, the term in parentheses is small. This result
reveals that when the state $\mathcal{A}$ approaches equilibrium,
$z_t$ is still very large and the HSC grows linearly. Further, we
notice that after the equilibrium, the HSC becomes stable and
the tip $z_t$ behaves as the order $\mathcal{O}(z_h)$.
Therefore, in these cases, the tip $z_t$ undergoes
discontinuous transition from the order $\mathcal{O}(l)$ to
the order $\mathcal{O}(z_h)$. In addition, when the size of the
subregion $\mathcal{A}$ goes to infinity, the HSC  persists
the linear growth all the way to the late time.

\subsubsection{Continuous transition at the late time}
In this stage, we mainly consider the case that the transition
of $z_t$ is continuous, namely, $z_t\rightarrow z_c$ for
$t\rightarrow t_s$. Then the constant $E(z_c,z_t)$ at $t_s$ is
\begin{equation}
  E=-\frac{g(z_c)}{2}\sqrt{\left(\frac{z_c}{z_c}\right)^{2d-2}-1}=0.
\end{equation}
Therefore, the equilibrium time $t_s$ and half-width $l$ can be
simplified as
\begin{align}
  &t_s=\int_0^{z_s}\frac{dz}{h(z)},\\
  l=\int_0^{z_s}&\frac{dz}{\sqrt{h(z)\left[\left(\frac{z_s}{z}\right)^{2d-2}-1\right]}},
\end{align}
where $z_s$ denotes that for $t=t_s$, we have $z_t=z_c=z_s$. The value of volume $\tilde{V}_{eq}$ at the equilibrium is given by
\begin{align}
  \tilde{V}_{eq}=&\int_0^{z_s}\frac{dz}{z^d}\sqrt{\frac{1}{h(z)}}\int_z^{z_s}\frac{dy}{\sqrt{h(y)\left(\frac{z_s^{2d-2}}{y^{2d-2}}-1\right)}}\nonumber\\
=&z_s^{2-d}\int_0^{1}\frac{dz}{z^d}\sqrt{\frac{1}{h(z_sz)}}\int_{z}^1\frac{dy}{\sqrt{h(z_sy)\left(y^{-2d+2}-1\right)}}.
\end{align}
In the limit of $l\gg z_h$, the tip $z_s\rightarrow z_h$. Thus one has
\begin{equation}
  z_s=z_h(1-\epsilon'), \quad\; \epsilon'\ll 1.
\end{equation}
In addition, near equilibrium we find both $z_t$ and $z_c$ approach $z_s$, which can be written as
\begin{align}
  z_t=z_s\left(1+\frac{\delta}{2d-2}\right),\\
  z_c=z_t\left(1-\frac{\epsilon^2}{2d-2}\right),
\end{align}
where $\delta$ and $\epsilon$ are both small parameters. According
to these relations, when $z_t$ approaches $z_c$, we can express
the constant $E(z_c,z_t)$ as
\begin{equation}
  E=\frac{-1}{2}g(z_t)\epsilon+\cdots.
\end{equation}

Next, we will analyze the relations between two parameters
$\epsilon$ and $\delta$. For $t\rightarrow t_s$, from $
l(t_s)=l(t)$ we have
\begin{equation}
  \int_0^{z_s}\frac{dz}{\sqrt{h(z)\left[\left(\frac{z_s}{z}\right)^{2d-2}-1\right]}}=\int_{z_c}^{z_t}\frac{dz}{\sqrt{\left(\frac{z_t}{z}\right)^{2d-2}-1}}+\int_{0}^{z_c}\frac{dz}{\sqrt{H(z)}}.
\end{equation}

Denoting
\begin{equation}
  F(z_s):=z_s\int_0^{1}\frac{dz}{\sqrt{h(z_sz)\left(z^{2-2d}-1\right)}},
\end{equation}
the above equation can be reduced to
\begin{equation}
  F'(z_s)\frac{-z_s\delta}{2d-2}+\cdots=\frac{g(z_t)}{h(z_t)}\frac{-z_t\epsilon}{2d-2}+\cdots.
\end{equation}

To the leading order, we obtain the relation between two parameters as
\begin{equation} \label{eq_de}
  \delta=\frac{g(z_s)}{h(z_s)F'(z_s)}\epsilon.
\end{equation}
With (\ref{eq_de}), we can express time $t-t_s$ by parameter
$\epsilon$, which is
\begin{equation}\label{eq_ts}
  t-t_s=\frac{g(z_s)z_s}{2}\left(\frac{1}{(d-1)h^2(z_s)F'(z_s)}-G(z_s)\right)\epsilon+\cdots=k\,\epsilon+\cdots,
\end{equation}
where
\begin{equation}
  k=\frac{g(z_s)z_s}{2}\left(\frac{1}{(d-1)h^2(z_s)F'(z_s)}-G(z_s)\right), \qquad G(z_s)=\int_0^1\frac{dy}{h(z_sy)}\frac{1}{\sqrt{h(z_sy)(y^{2-2d}-1)}}.
\end{equation}

Next we turn to evaluate the difference $\tilde{V}(t)-\tilde{V}_{eq}$ near $t=t_s$, which is
\begin{equation}\label{eq_saV}
  \tilde{V}(t)-\tilde{V}_{eq}=\tilde{V}_{AdS}(t)+N(z_c)-M(z_c)+M(z_c)-I(z_s),
\end{equation}
where we denote
\begin{align}
  N(z_c):=&\int_0^{z_c}\frac{dz}{z^d}\sqrt{\frac{1}{h(z)}\left(\frac{-E^2}{H(z)}+1\right)}\left(l-\int_0^z\frac{dy}{\sqrt{H(y)}}\right),\\
  M(z_c):=&\int_0^{z_c}\frac{dz}{z^d}\sqrt{\frac{1}{h(z)}}\int_z^{z_t}\frac{dy}{\sqrt{h(y)\left(\frac{z_t^{2d-2}}{y^{2d-2}}-1\right)}},\\
  I(z_s):=&z_s^{2-d}\int_0^{1}\frac{dz}{z^d}\sqrt{\frac{1}{h(z_sz)}}\int_{z}^1\frac{dy}{\sqrt{h(z_sy)\left(y^{-2d+2}-1\right)}}.
\end{align}

As for the first term in (\ref{eq_saV}), we find
\begin{equation}
  \tilde{V}_{AdS}(t)=\frac{1}{z_t^{d-2}}\frac{\epsilon^3}{3(d-1)^2}+\cdots.
\end{equation}

Since the UV divergence can be subtracted by $\tilde{V}_{vac}$,
the difference between integrands is only significant near
$z=z_c$. As for the term $N(z_c)-M(z_c)$, we find
\begin{equation}
  N(z_c)-M(z_c)=\frac{-1}{z_c^{d-2}}\frac{1}{h(z_c)}\frac{\epsilon}{d-1}+\cdots.
\end{equation}
As for the term $M(z_c)-I(z_s)$, we find
\begin{equation}
  M(z_c)-I(z_s)=\frac{1}{2(d-1)}\frac{I'(z_s)g(z_s)z_s}{F'(z_s)h(z_s)}\epsilon+\cdots.
\end{equation}

Finally, we obtain the evolution behavior of subtracted volume $\Delta\tilde{V}(t)-\Delta\tilde{V}_{eq}$ near $t=t_s$, where the symbol ``$\Delta$'' refers to subtracting the vaccum value $\tilde{V}_{vac}$. Then the expression gives
\begin{align}
  \Delta\tilde{V}(t)-\Delta\tilde{V}_{eq}=&\frac{z_s}{(d-1)h(z_s)}\left[\frac{1}{2}\frac{I'(z_s)}{F'(z_s)}g(z_s)-\frac{1}{z_s^{d-1}}\right]\epsilon+\cdots\label{eq_sv}\\
  =&C_s\left(t-t_s\right)+\cdots,
\end{align}  with
\begin{align}C_s:=&\frac{h(z_s)}{z_s^{d-1}g(z_s)}\frac{I'(z_s)z_s^{d-1}g(z_s)-2F'(z_s)}{1-(d-1)G(z_s)F'(z_s)h^2(z_s)}.
 \end{align}
The sign of coefficient $C_s$ is not manifest. Firstly, in
(\ref{eq_sv}) we find two equations, which are
\begin{equation}
  I'(z_s)=\lim\limits_{\theta \rightarrow 0}\frac{1}{z_s^{d-1}}\int_0^{1}\frac{dz}{z^d}\sqrt{\frac{1}{h(z_s z)}}\Bigg(\frac{1}{\sqrt{h(z_s)2(d-1)\theta}}-\int_z^{1-\theta}\frac{(d-1)y^{d-1}dy}{\sqrt{h(z_s y)}(1-y^{2d-2})^{3/2}}\Bigg)
\end{equation}
and
\begin{equation}
  F'(z_s)=\lim\limits_{\theta \rightarrow 0}\Bigg(\frac{1}{\sqrt{h(z_s)2(d-1)\theta}}-\int_0^{1-\theta}\frac{(d-1)y^{d-1}dy}{\sqrt{h(z_s y)}(1-y^{2d-2})^{3/2}}\Bigg).
\end{equation}

Since the above equations are divergent, we should take the limit
carefully. Fortunately, the divergence can be eliminated when
computing
\begin{equation}
  \frac{I'(z_s)}{F'(z_s)}>\frac{1}{z_s^{d-1}}\int_0^1\frac{dz}{z^d}\sqrt{\frac{1}{h(z_s\,z)}}\gg \frac{1}{z_s^{d-1}}.
\end{equation}

Therefore, it is easy to judge the coefficient in ({\ref{eq_sv}})
is positive. Further, from  (\ref{eq_ts}) we know that for a
continuous transition, $k$ is negative in
general\footnote{Actually, for $d=2$ SAdS case, $k=0$ and $t-t_s$
is determined by the term $\mathcal{O}(\epsilon^2)$. One can
check that the coefficient is still negative.}. Therefore, $C_s$
is negative. As a result, we conclude that for a continuous
transition, the subtracted volume
$\Delta\tilde{V}(t)-\Delta\tilde{V}_{eq}$ decreases linearly near
equilibrium.

This result is again in contrast to the one in CV conjecture. Perhaps the difference could be understood from the holographic point of view. In CV conjecture, the extremal surface is always a Cauchy surface during the evolution, while in
the subregion conjecture, the extremal surface $\Gamma_{\mathcal{A}}$ is always reaching an equilibrium configuration at the late time, even in the large size limit.

\section{Conclusions and discussions}\label{sec_Con}
In this paper, we have investigated the complexity of a mixed state with a gravitational dual in strongly coupled field theory. Specifically, we have
analytically evaluated the HSC over a general Vaidya-AdS
spacetime, where the specific form of the metric is not necessary.
Based on the strategy presented in \cite{Liu:2013qca}, we have
found three characteristic stages during the evolution of HSC. At
the very early time right after the null shell begins to fall
down, the HSC grows with a linear manner no matter the size of the
subregion is. Since UV region is thermalized prior to IR
region in this setting, the result reveals that the change of
complexity is sensitive to the UV modes of
different systems. While at the intermediate stage when
$\Gamma_{\mathcal{A}}$ is close to the critical configuration, the
HSC also exhibits a linear growth in the large size limit,
while the rate of growth is sensitive to the spacetime dimension
as well as the type of quench. At the late time,
if the transition is discontinuous, then the HSC increases
linearly which is similar to the result in the intermediate stage,
otherwise, the HSC drops down continuously to a stable value.
The evolution behavior in two stages are very different from
the one obtained by CV conjecture.

We also compare the growth rates of HSC with the Lloyd bound in
the SAdS cases. We find that with some choices of certain
parameter, the Lloyd bound is always saturated at the early time,
while at the intermediate stage, the growth rate is always less
than the Lloyd bound.

The discrepancy can be understood as follows. From
the gravity side, the reason leading to such a discrepancy seems evident, since in the CV conjecture, the surface with maximal
volume is always a Cauchy surface during the quench process, while
in the subregion conjecture, the tip $z_t$ of the extremal surface
$\Gamma_{\mathcal{A}}$ always decreases either continuously
or discontinuously to an equilibrium value $z_s$, even in the
large size limit. Therefore, the configurations of the extremal
surface as well as the rates of growth are generally different in these two conjectures. From the field theory side, when the subregion is large enough to
cover the whole boundary, the Hilbert spaces of these two states coincide. Therefore, the discrepancy may be caused by the different choices of the reference state and these phenomena may give some hints for determining the reference states in
 subregion conjectures.

One challenge is to understand the
drop of  complexity at the late time stage of evolution, which
seems to be a quite general phenomenon for HSC with finite size. For instance, in the
$(2+1)$-dimensional SAdS case, the complexity can be reduced
continuously to a lower level. However, as mentioned in literature, a process with decreasing complexity should be unstable from the
thermodynamical point of view \cite{Stanford:2014jda,Susskind:2014jwa,Susskind:2013lpa,Brown:2017jil}. This seems paradoxical and it is
very desirable to have a better understanding on the nature of holographic subregion complexity.

The work can be generalized directly to more general cases to describe some spatially homogenous and isotropic equilibrium processes as mentioned in \cite{Liu:2013qca}. Then the metric can be expressed as
\begin{equation}
  ds^2=\frac{1}{z^2}\left(-f(v,z)dv^2-2q(v,z)dvdz+\sum_{i=1}^{d-1}dx_i^2\right)
\end{equation}
where
\begin{equation}
  f(v,z):=1-\theta(v)g(z),\qquad q(v,z):=1-\theta(v)k(z).
\end{equation}

It is also desirable to explore the evolution of HSC under the CA
conjecture, since there is no ambiguity of choosing the arbitrary
length scale in the CA conjecture and the growth rate of
complexity is naturally related to the Lloyd bound.

As we  know, it is still an open problem to explore the dynamical behavior of  the quantum complexity in a strongly correlated system. Holography provides a plausible way to  examine the growth rate of complexity during the quench process. It is quite  intriguing but challenging to disclose more properties of  HSC  in other sorts of dynamical process  and compare its dynamical behavior with that in other systems such as  the circuit complexity.

Finally, it is interesting to investigate the dynamical behavior of some new quantities which are analogous to the notions in the context of holographic entanglement entropy, such as mutual complexity \cite{Alishahiha:2018lfv} and complexity purification \cite{Agon:2018zso,Caceres:2018blh,Camargo:2018eof,Ghodrati:2019hnn}.

\section*{Acknowledgments}
We are very grateful to Qinghua Zhu, Zhibin Li,
Zhuoyu Xian, Jing Gao for helpful discussions and suggestions.
This work is supported by the Natural Science Foundation of China
under Grant No. 11575195, 11805083 and 11875053. Y.L. also acknowledges the
support from 555 talent project of Jiangxi Province.


\begin{thebibliography}{10}
    \bibitem{Stanford:2014jda}
  D.~Stanford and L.~Susskind,
  ``Complexity and Shock Wave Geometries,''
  Phys.\ Rev.\ D {\bf 90}, no. 12, 126007 (2014)
  [arXiv:1406.2678 [hep-th]].

  \bibitem{Brown:2015bva}
  A.~R.~Brown, D.~A.~Roberts, L.~Susskind, B.~Swingle and Y.~Zhao,
  ``Holographic Complexity Equals Bulk Action?,''
  Phys.\ Rev.\ Lett.\  {\bf 116}, no. 19, 191301 (2016)
  [arXiv:1509.07876 [hep-th]].

  \bibitem{Alishahiha:2015rta}
  M.~Alishahiha,
  ``Holographic Complexity,''
  Phys.\ Rev.\ D {\bf 92}, no. 12, 126009 (2015)
  [arXiv:1509.06614 [hep-th]].

  \bibitem{Carmi:2016wjl}
  D.~Carmi, R.~C.~Myers and P.~Rath,
  ``Comments on Holographic Complexity,''
  JHEP {\bf 1703}, 118 (2017)
  [arXiv:1612.00433 [hep-th]].

  \bibitem{Bakhshaei:2017qud} 
  E.~Bakhshaei, A.~Mollabashi and A.~Shirzad,
  ``Holographic Subregion Complexity for Singular Surfaces,''
  Eur.\ Phys.\ J.\ C {\bf 77}, no. 10, 665 (2017)
  [arXiv:1703.03469 [hep-th]].

  \bibitem{Lezgi:2019fqu} 
  M.~Lezgi and M.~Ali-Akbari,
  ``A note on holographic subregion complexity and QCD phase transition,''
  arXiv:1908.01303 [hep-th].

  \bibitem{Bhattacharya:2019zkb} 
  A.~Bhattacharya, K.~T.~Grosvenor and S.~Roy,
  ``Entanglement Entropy and Subregion Complexity in Thermal Perturbations around Pure-AdS,''
  arXiv:1905.02220 [hep-th].

  \bibitem{Zhou:2019jlh} 
  Y.~T.~Zhou, M.~Ghodrati, X.~M.~Kuang and J.~P.~Wu,
  ``Evolutions of entanglement and complexity after a thermal quench in massive gravity theory,''
  Phys.\ Rev.\ D {\bf 100}, no. 6, 066003 (2019)
  [arXiv:1907.08453 [hep-th]].

  \bibitem{Zhang:2019vgl} 
  S.~J.~Zhang,
  ``Subregion complexity in holographic thermalization with dS boundary,''
  Eur.\ Phys.\ J.\ C {\bf 79}, no. 8, 715 (2019)
  [arXiv:1905.10605 [hep-th]].

  \bibitem{Fujita:2018xkl} 
  M.~Fujita,
  ``Holographic subregion complexity of a (1+1)-dimensional $p$-wave superconductor,''
  PTEP {\bf 2019}, no. 6, 063B04 (2019)
  [arXiv:1810.09659 [hep-th]].

  \bibitem{Karar:2019wjb} 
  S.~Karar, R.~Mishra and S.~Gangopadhyay,
  ``Holographic complexity of boosted black brane and Fisher information,''
  Phys.\ Rev.\ D {\bf 100}, no. 2, 026006 (2019)
  [arXiv:1904.13090 [hep-th]].

  \bibitem{Auzzi:2019fnp} 
  R.~Auzzi, S.~Baiguera, A.~Mitra, G.~Nardelli and N.~Zenoni,
  ``Subsystem complexity in warped AdS,''
  JHEP {\bf 1909}, 114 (2019)
  [arXiv:1906.09345 [hep-th]].

  \bibitem{Ghosh:2019jgd} 
  A.~Ghosh and R.~Mishra,
  ``Inhomogeneous Jacobi equation and Holographic subregion complexity,''
  arXiv:1907.11757 [hep-th].

  \bibitem{Auzzi:2019mah} 
  R.~Auzzi, G.~Nardelli, F.~I.~Schaposnik Massolo, G.~Tallarita and N.~Zenoni,
  ``On volume subregion complexity in Vaidya spacetime,''
  arXiv:1908.10832 [hep-th].

  \bibitem{Braccia:2019xxi} 
  P.~Braccia, A.~L.~Cotrone and E.~Tonni,
  ``Complexity in the presence of a boundary,''
  arXiv:1910.03489 [hep-th].

  \bibitem{Agon:2018zso}
  C.~A.~Agón, M.~Headrick and B.~Swingle,
  ``Subsystem Complexity and Holography,''
  arXiv:1804.01561 [hep-th].

  \bibitem{Caceres:2018luq}
  E.~Caceres and M.~L.~Xiao,
  ``Complexity-action of subregions with corners,''
  arXiv:1809.09356 [hep-th].

  \bibitem{Alishahiha:2018lfv}
  M.~Alishahiha, K.~Babaei Velni and M.~R.~Mohammadi Mozaffar,
  ``Subregion Action and Complexity,''
  arXiv:1809.06031 [hep-th].

  \bibitem{Ben-Ami:2016qex}
  O.~Ben-Ami and D.~Carmi,
  ``On Volumes of Subregions in Holography and Complexity,''
  JHEP {\bf 1611}, 129 (2016)
  [arXiv:1609.02514 [hep-th]].

  \bibitem{Abt:2018ywl}
  R.~Abt, J.~Erdmenger, M.~Gerbershagen, C.~M.~Melby-Thompson and C.~Northe,
  ``Holographic Subregion Complexity from Kinematic Space,''
  arXiv:1805.10298 [hep-th].

  \bibitem{Du:2018uua}
  L.~P.~Du, S.~F.~Wu and H.~B.~Zeng,
  ``Holographic complexity of the disk subregion in (2+1)-dimensional gapped systems,''
  arXiv:1803.08627 [hep-th].

  \bibitem{Abt:2017pmf}
  R.~Abt, J.~Erdmenger, H.~Hinrichsen, C.~M.~Melby-Thompson, R.~Meyer, C.~Northe and I.~A.~Reyes,
  ``Topological Complexity in AdS3/CFT2,''
  arXiv:1710.01327 [hep-th].

  \bibitem{Roy:2017uar}
  P.~Roy and T.~Sarkar,
  ``Subregion holographic complexity and renormalization group flows,''
  Phys.\ Rev.\ D {\bf 97}, no. 8, 086018 (2018)
  [arXiv:1708.05313 [hep-th]].

  \bibitem{Banerjee:2017qti}
  S.~Banerjee, J.~Erdmenger and D.~Sarkar,
  ``Connecting Fisher information to bulk entanglement in holography,''
  JHEP {\bf 1808}, 001 (2018)
  [arXiv:1701.02319 [hep-th]].

  \bibitem{Zangeneh:2017tub}
  M.~Kord Zangeneh, Y.~C.~Ong and B.~Wang,
  ``Entanglement Entropy and Complexity for One-Dimensional Holographic Superconductors,''
  Phys.\ Lett.\ B {\bf 771}, 235 (2017)
  [arXiv:1704.00557 [hep-th]].

  \bibitem{Bhattacharya:2018oeq}
  A.~Bhattacharya and S.~Roy,
  ``Holographic Entanglement Entropy, Subregion Complexity and Fisher Information metric of 'black' Non-SUSY D3 Brane,''
  arXiv:1807.06361 [hep-th]

  \bibitem{Zhang:2018qnt}
  S.~J.~Zhang,
  ``Subregion complexity and confinement-deconfinement transition in a holographic QCD model,''
  arXiv:1808.08719 [hep-th].

  \bibitem{Roy:2017kha}
  P.~Roy and T.~Sarkar,
  ``Note on subregion holographic complexity,''
  Phys.\ Rev.\ D {\bf 96}, no. 2, 026022 (2017)
  [arXiv:1701.05489 [hep-th]].

  \bibitem{Chen:2018mcc}
  B.~Chen, W.~M.~Li, R.~Q.~Yang, C.~Y.~Zhang and S.~J.~Zhang,
  ``Holographic subregion complexity under a thermal quench,''
  JHEP {\bf 1807}, 034 (2018)
  [arXiv:1803.06680 [hep-th]].

  \bibitem{Ling:2018xpc}
  Y.~Ling, Y.~Liu and C.~Y.~Zhang,
  ``Holographic Subregion Complexity in Einstein-Born-Infeld theory,''
  arXiv:1808.10169 [hep-th].
  
  \bibitem{Liu:2013qca}
  H.~Liu and S.~J.~Suh,
  ``Entanglement growth during thermalization in holographic systems,''
  Phys.\ Rev.\ D {\bf 89}, no. 6, 066012 (2014)
  [arXiv:1311.1200 [hep-th]].

\bibitem{Balasubramanian:2010ce}
  V.~Balasubramanian {\it et al.},
  ``Thermalization of Strongly Coupled Field Theories,''
  Phys.\ Rev.\ Lett.\  {\bf 106}, 191601 (2011)
  [arXiv:1012.4753 [hep-th]].

  \bibitem{Balasubramanian:2011ur}
  V.~Balasubramanian {\it et al.},
  ``Holographic Thermalization,''
  Phys.\ Rev.\ D {\bf 84}, 026010 (2011)
  [arXiv:1103.2683 [hep-th]].

  \bibitem{Chapman:2018dem}
  S.~Chapman, H.~Marrochio and R.~C.~Myers,
  ``Holographic complexity in Vaidya spacetimes. Part I,''
  JHEP {\bf 1806}, 046 (2018)
  [arXiv:1804.07410 [hep-th]].

  \bibitem{Kim:2017qrq}
  R.~Q.~Yang, C.~Niu, C.~Y.~Zhang and K.~Y.~Kim,
  ``Comparison of holographic and field theoretic complexities for time dependent thermofield double states,''
  JHEP {\bf 1802}, 082 (2018)
  [arXiv:1710.00600 [hep-th]].

  \bibitem{Susskind:2014jwa}
  L.~Susskind and Y.~Zhao,
  ``Switchbacks and the Bridge to Nowhere,''
  arXiv:1408.2823 [hep-th].

  \bibitem{Susskind:2013lpa}
  L.~Susskind,
  ``New Concepts for Old Black Holes,''
  arXiv:1311.3335 [hep-th].

  \bibitem{Brown:2017jil}
  A.~R.~Brown and L.~Susskind,
  ``Second law of quantum complexity,''
  Phys.\ Rev.\ D {\bf 97}, no. 8, 086015 (2018)
  doi:10.1103/PhysRevD.97.086015
  [arXiv:1701.01107 [hep-th]].

  \bibitem{Caceres:2018blh}
  E.~Cáceres, J.~Couch, S.~Eccles and W.~Fischler,
  ``Holographic Purification Complexity,''
  arXiv:1811.10650 [hep-th].

  \bibitem{Camargo:2018eof}
  H.~A.~Camargo, P.~Caputa, D.~Das, M.~P.~Heller and R.~Jefferson,
  ``Complexity as a novel probe of quantum quenches: universal scalings and purifications,''
  arXiv:1807.07075 [hep-th].

  \bibitem{Ghodrati:2019hnn}
  M.~Ghodrati, X.~M.~Kuang, B.~Wang, C.~Y.~Zhang and Y.~T.~Zhou,
  ``The connection between holographic entanglement and complexity of purification,''
  arXiv:1902.02475 [hep-th].
\end{thebibliography}
\end{document}